\documentstyle[epsf,11pt]{article}

\newcommand{\lsim}{~\ooalign{\hfil{\mbox{\Large${}_{\sim}$}}\hfil\crcr{\mbox{\raisebox{0.4ex}{$<$}}}}~}

\makeatletter
\@addtoreset{equation}{section}
\makeatother

\begin{document}
\begin{titlepage}
\begin{flushright} NUP-A-2000-14\\
hep-ph/0007225
\end{flushright}
\vspace{1cm}
\begin{center}
{\Large{\bf Compositeness Condition and Vacuum Stability}}\\
\vspace{1.5cm}
Eizou UMEZAWA\footnote{e-mail: umezawa@phys.cst.nihon-u.ac.jp}\\
\vspace{0.2cm}
{\it Department of Physics, College of Science and Technology,\\
Nihon University, Tokyo 101-8308}
\end{center}
\vspace{3.5cm}
\begin{abstract}
We consider what occurs when we remove one of the compositeness conditions proposed by Bardeen, Hill and Lindner that leads to predictions for the top quark mass conflicting with the experimental value. Through this consideration the condition for the Higgs particle to be the composite particle is reconsidered. We show that in this case, (I) the Higgs-Yukawa system of the standard model becomes equivalent to a non-local four-fermi system at a high-energy scale $\Lambda$, (II) The Higgs-Yukawa sector of the model becomes useless above the scale because the vacuum state cannot be defined. We regard the two phenomena as indications of the compositeness of the Higgs particle. It is suggested that the new physics above $\Lambda$ contains bi-local fields.
\end{abstract}
\end{titlepage}

\section{Introduction}
The top quark condensation is an attractive idea that explains the electroweak symmetry breaking in the absence of fundamental scalar bosons, and gives an understanding of that the top quark mass is of the order of the electroweak scale~\cite{T-condensate model,BHL}. Bardeen, Hill and Lindner (BHL) have proposed an interesting scenario for the top condensation~\cite{BHL}. In their model, the renormalization group (RG) equations play an important role, and the composite nature of the Higgs particle is reflected in the boundary conditions for the coupling constants of the standard model, namely, the compositeness conditions of BHL.

The usual standard model is considered to well describe physical phenomena in the low energies where the Higgs particle can be regarded as elementary particle. On the other hand, if the Higgs particle is composed of some elementary particle, the Higgs-Yukawa sector of the model will be useless in some high-energy region because the lagrangian written in terms of the local Higgs field must be useless to describe the inside of the Higgs particle. In the scenario of BHL, the Higgs-Yukawa sector becomes useless at a high-energy scale $\Lambda$ and above owing to the divergence of the Yukawa coupling constant for the top quark at the scale. Besides, BHL have also supposed that the Higgs field becomes a non-propagating auxiliary field at $\Lambda$, and so the Higgs-Yukawa system is equivalent to a local four-fermi system at the scale. These two phenomena can be considered to indicate the compositeness of the Higgs particle.

Using one of the compositeness conditions of BHL, one can predict the top quark mass depending on the value of $\Lambda$. Unfortunately, the predicted masses are larger that the experimental value as for $\Lambda \le m_{Pl}$ ($m_{Pl} \simeq 10^{19}$ GeV is Planck scale), and several improvements and extensions of BHL's model have been studied~\cite{improvement}.

Our attempt to improve the model is very simple: we merely remove the condition that leads to the predictions contradicted by the experiment from the compositeness conditions. We consider what occurs in this case, and reconsider the condition for the Higgs particle to be the composite particle.

In Refs.~\cite{BKSWS,BKMN}, it has been shown that if we use the $\beta$ functions of the mass independent renormalization scheme in BHL's model, we cannot properly describe the spontaneous symmetry breaking. This is the same in our scenario for the composite Higgs particle (See Appendix A). Ideally, we should use Wilson's renormalization scheme~\cite{Wilson} as discussed in Refs.~\cite{BKSWS,BKMN}, however, the treatment of the RG equation is not so easy in this scheme. Actually, BHL have used the $\beta$ functions of the mass independent renormalization scheme in the continuous theory instead of ones of Wilsonian's. The validity of this substitution has been discussed in Ref.~\cite{BKMN} in leading order of the $1/N_{c}$ expansion, where $N_{c}$ is the number of colors\footnote{See also Refs.~\cite{BKSWS,HKKN}}. There, it was concluded that this substitution is valid as an approximation except for the $\beta$ function for the mass parameter. In this paper, we will also define the renormalization transformation in our cutoff model with having the Wilson renormalization approach in mind, and give the $\beta$ functions for the quartic ($\lambda$) and the quadratic ($m^{2}$) coupling constants of the Higgs self-interactions. To give them, we will use the perturbative expansion with respect to all the coupling constants including the Yukawa coupling constant for the top quark, which does not diverge at $\Lambda$ in our scenario. We will see that our results coincide with the statement of Ref.~\cite{BKMN}. On the basis of this observation, disregarded the problem of the gauge invariance (See Appendix A), we use the $\beta$ functions of the ${\overline {\rm MS}}$ scheme for all the dimensionless coupling constants in this paper except for a few cases.

In the next section, we review the compositeness conditions of BHL. In \S 3, we attempt to remove the condition that leads to the quark mass predictions contradicted by the experiment. The condition for the Higgs particle to be the composite particle is reconsidered in this section. In \S 4, predictions for the Higgs boson mass are given. Section 5 is devoted to summary and discussions. In Appendix A, we define the renormalization transformation and give the $\beta$ functions for $\lambda$ and $m^{2}$ with having Wilson's renormalization approach in mind.
\section{Compositeness conditions of BHL}
We have Wilson's renormalization approach in mind. The effective lagrangian density of the standard model at an energy scale $\mu$ will be
\begin{equation}
{\cal L}={\cal L}_{qg}+\left(D_{\mu}\Phi\right)^{\dag}D^{\mu}\Phi-m^{2}\Phi^{\dag}\Phi-\lambda\bigl(\Phi^{\dag}\Phi\bigr)^{2}+{\cal L}_{Y}+{\cal L}^{{\rm H.D.}},\label{lagrangian}
\end{equation}
where
${\cal L}_{qg}$ denotes the usual gauge invariant lagrangian for the gauge and the fermion fields, $\Phi$ is the Higgs doublet, $D_{\mu}$ is the gauge covariant derivative and ${\cal L}^{{\rm H.D.}}$ denotes the higher dimensional terms. The Yukawa interaction terms are
\begin{equation}
{\cal L}_{Y}=-\sum_{i}^{1,2,3}\left\{y_{i}^{d}\left({\overline Q}_{i}^{d}\Phi q_{i}^{d}+\mbox{h.c.}\right)+y_{i}^{u}\left({\overline Q}_{i}^{u}\widetilde{\Phi} q_{i}^{u}+\mbox{h.c.}\right)+y_{i}^{l}\left({\overline L}_{i}\Phi l_{i}+\mbox{h.c.}\right)\right\},\label{Yukawa}
\end{equation}
where $i=1,2,3$ denote the generations, $\widetilde{\Phi}=i\sigma_{2}\Phi^{\ast}$, $Q_{i}^{u(d)}$ are the left-handed quark doublets in which the up (down) type components are mass eigen states, $q_{i}^{u(d)}$ are the right-handed up (down) type quark singlets, $L_i$ are the left-handed lepton doublets and $l_{i}$ are the right-handed lepton singlets.

Now suppose that the running coupling constants  behave as\footnote{This explanation of BHL's scenario is in accord with Ref.~\cite{BKSWS}.}
\begin{equation}
\lambda_{n}(\mu)=Z^{-n/2}(\mu)\widehat{\lambda}_{n}
\end{equation}
near $\mu \lsim \Lambda$, where $\lambda_{n}(\mu)$ denotes the coupling constant of the interaction term that is proportional to the power of $n$ of $\Phi$, $\widehat{\lambda}_{n}$ is a finite constant, and the small factor $Z(\mu)$ vanishes at a high-energy scale $\Lambda$:
\begin{equation}
\lim_{\mu \to \Lambda}Z(\mu)=0.\label{condition for Z}
\end{equation}
Specifically, we write as
\begin{equation}
y_{i}^{u,d,l}(\mu)=Z^{-1/2}(\mu)\widehat{y}^{u,d,l}_{i},~~\lambda(\mu)=Z^{-2}(\mu)\widehat{\lambda},~~m^{2}(\mu)=Z^{-1}(\mu)\widehat{m}^{2},\label{compositeness0-1}
\end{equation}
where $\widehat{y}^{u,d,l}_{i}$, $\widehat{\lambda}$ and $\widehat{m}^{2}$ are finite constants ($\widehat{y}^{u}_{3},~\widehat{m}^{2} \ne 0$).
In this case, we have
\begin{equation}
{\cal L}={\cal L}_{qg}+Z(\mu)\bigl(D_{\mu}\widehat{\Phi}\bigr)^{\dag}D^{\mu}\widehat{\Phi}-\widehat{m}^2\widehat{\Phi}^{\dag}\widehat{\Phi}-\widehat{\lambda}\bigl(\widehat{\Phi}^{\dag}\widehat{\Phi}\bigr)^{2}+\widehat{{\cal L}}_{Y}+\widehat{{\cal L}}^{{\rm H.D.}}
\end{equation}
near $\mu \lsim \Lambda$, where $\widehat{\Phi}=\Phi/Z^{1/2}(\mu)$, $\widehat{{\cal L}}_{Y}$ is given by replacing $y_{i}^{u,d,l}$ and $\Phi$ in Eq.~(\ref{Yukawa}) with $\widehat{y}_{i}^{u,d,l}$ and $\widehat{\Phi}$, respectively, and $\widehat{{\cal L}}^{{\rm H.D.}}$ is given by replacing $\lambda_{n}(\mu)$ and $\Phi$ in ${\cal L}^{{\rm H.D.}}$ with $\widehat{\lambda}_{n}$ and $\widehat{\Phi}$, respectively. Further, if
\begin{eqnarray}
\widehat{\lambda}&=&0,\label{compositeness0-2}\\
(\widehat{\lambda}_{n} \mbox{ in } \widehat{{\cal L}}^{{\rm H.D.}})&=&0,\label{compositeness0-3}
\end{eqnarray}
the lagrangian reduces to
\begin{equation}
{\cal L}={\cal L}_{qg}-\widehat{m}^{2}\widehat{\Phi}^{\dag}\widehat{\Phi}-\widehat{y}^{u}_{3}\bigl({\overline Q}_{3}^{u}\widetilde{\widehat{\Phi}} q_{3}^{u}+\mbox{h.c.}\bigr)\label{reduce}
\end{equation}
for $\mu \to \Lambda$, where we have considered that $\widehat{y}^{u,d,l}_{i=1,2}$ and $\widehat{y}^{d,l}_{3}$ vanish or become negligible compared to $\widehat{y}^{u}_{3}$. Through the auxiliary field method, we can show that the system of Eq.~(\ref{reduce}) is equivalent to one of the four-fermi interaction supplemented with gauge interaction,
\begin{equation}
{\cal L}'={\cal L}_{qg}+G\left({\overline Q_{3}^{u}}q_{3}^{u}\right)\left({\overline q_{3}^{u}}Q_{3}^{u}\right),~~~ G=\frac{\left(\widehat{y}^{u}_{3}\right)^{2}}{\widehat{m}^{2}}.
\end{equation}
Therefore, we can describe the Higgs particle by alternative lagrangian below $\Lambda$ : one is the Higgs-Yukawa lagrangian of the standard model, and another is the lagrangian for the fermions that has the local four-fermi interaction term with other higher dimensional terms generated through the renormalization transformation from $\Lambda$. On the other hand, at $\Lambda$ and above, we cannot use the Higgs-Yukawa lagrangian as follows. From Eqs.~(\ref{compositeness0-1}) and (\ref{compositeness0-2}), we have
\begin{eqnarray}
\lim_{\mu \to \Lambda}\frac{1}{\left\{y_{3}^{u}(\mu)\right\}^2}&=&0,\label{composite condition1}\\
\lim_{\mu \to \Lambda}\frac{\lambda(\mu)}{\left\{y_{3}^{u}(\mu)\right\}^4}&=&0,\label{composite condition2}
\end{eqnarray}
which are the compositeness conditions for the running coupling constants proposed by BHL. The condition (\ref{composite condition1}) means that the Higgs-Yukawa sector of the standard model becomes useless at $\Lambda$ and above.

Here, we arrange the phenomena arising from the compositeness conditions of BHL.

\begin{enumerate}
\renewcommand{\labelenumi}{(\roman{enumi})}
 \item the Higgs-Yukawa system of the standard model becomes equivalent to the local four-fermi interaction system at $\Lambda$.

 \item the Higgs-Yukawa sector of the standard model becomes useless at $\Lambda$ and above because the Yukawa coupling constant for the top quark diverges at the scale.
\end{enumerate}

\noindent
The two phenomena can be considered to indicate the compositeness of the Higgs particle.

Using the condition (\ref{composite condition1}), we can predict the top quark mass depending on $\Lambda$. We can solve\footnote{Practically, one may solve the differential equation for $1/\alpha_{t}$.}
\begin{equation}
\frac{d\alpha_{t}}{dt}=\beta_{\alpha_{t}},~~\alpha_{t}=\frac{\left(y_{3}^{u}\right)^2}{4\pi}\label{RGE-at}
\end{equation}
using the condition as a boundary condition, where $t=\ln(\mu/m_{Z}),~m_{Z}=91.187 \mbox{~GeV}$ and the 1-loop $\beta$ function for $\alpha_{t}$ is
\begin{equation}
4\pi \beta_{\alpha_{t}}=\alpha_{t}\left(9\alpha_{t}-\frac{17}{10}\alpha_{1}-\frac{9}{2}\alpha_{2}-16\alpha_{3}\right).
\end{equation}
The gauge coupling constants
\begin{equation}
\alpha_{i}=\frac{g_{i}^{2}}{4\pi}~\left(i=1,2,3,~~g_{1}=\sqrt{\frac{5}{3}}g'\right)
\end{equation}
obtained from 1-loop $\beta$ functions are
\begin{eqnarray}
\alpha_{1}^{-1}(t)&=&\alpha_{1}^{-1}(t=0)-\frac{41}{20\pi}t,~~~\alpha_{1}(t=0)=0.016953,\\
\alpha_{2}^{-1}(t)&=&\alpha_{2}^{-1}(t=0)+\frac{19}{12\pi}t,~~~\alpha_{2}(t=0)=0.033817,\\
\alpha_{3}^{-1}(t)&=&\alpha_{3}^{-1}(t=0)+\frac{7}{2\pi}t,~~~\alpha_{3}(t=0)=0,119.
\end{eqnarray}
All the $\beta$ functions used above are of the ${\overline {\rm MS}}$ scheme in the continuous theory. Because $\alpha_{t}(\mu)$ have been determined thanks to the condition (\ref{composite condition1}), the top quark mass is given by
\begin{equation}
{\overline m_{t}}(\mu)=\sqrt{2\pi\alpha_{t}(\mu)}\,v,~~v=246.22 \mbox{~GeV},\label{v}
\end{equation}
where $v$ is the vacuum expectation value of the Higgs field.

\begin{figure}[t]
 \leavevmode
 \epsfxsize= 100mm
 \epsfysize= 70mm
 \centerline{\epsfbox{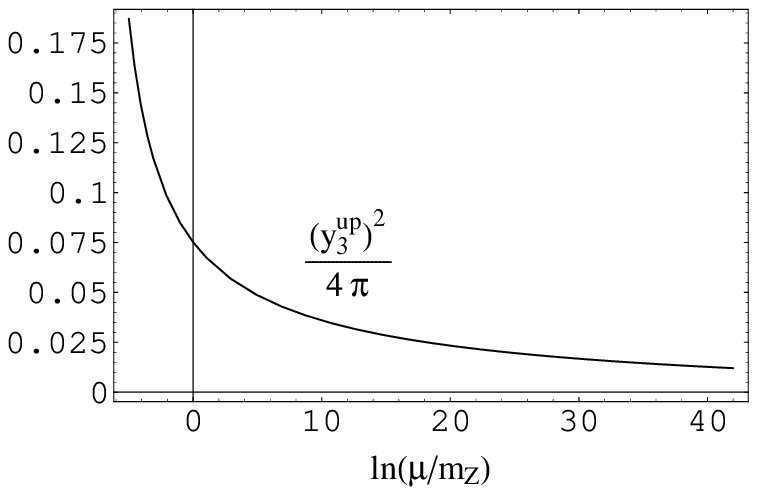}} 
 \caption{The RG evolution of $\alpha_{t}(\mu)=\left\{y_{3}^{u}(\mu)\right\}^{2}/(4\pi)$ for the boundary condition $\alpha_{t}(m_{Z})=0.075168$ obtained from $m_{t}=173.8$ GeV (${\overline m_{t}}({\overline m_{t}})=163.9$ GeV).}
 \label{at}
\end{figure}

Unfortunately, the results of the top quark mass given in this way are larger than the experimental value, $m_{t}=173.8 \pm 5.2$ GeV\footnote{The experimental values used in this paper are taken from Ref.~\cite{PDG}.}. Actually, if we determine $y_{3}^{u}(\mu)$ by solving Eq.~(\ref{RGE-at}) with the boundary condition obtained from the experimental value of the top quark mass, the condition (\ref{composite condition1}) does not hold as for $\Lambda \le m_{p}$. For example, we give the running of $\alpha_{t}$ for $\Lambda=10^{4}$ GeV and $m_{t}=173.8$ GeV in Fig.~\ref{at}. To this calculation we have converted the pole mass $m_{t}=173.8$ GeV into the ${\overline {\rm MS}}$ mass ${\overline m_{t}}({\overline m_{t}})=163.9$ GeV using
\begin{eqnarray}
&&{\overline m_{t}}({\overline m_{t}})=m_{t}(1+\delta)^{-1}\equiv {\overline m_{t}},\label{pole}\\
&&\delta=\frac{4}{3}a_{3}+8.2366a_{3}^{2}+73.638a_{3}^{3}~~\left(a_{3}=\frac{\alpha_{3}({\overline m_{t}})}{\pi}\right)\nonumber
\end{eqnarray}
according to Ref.~\cite{pole}.
\section{Modification of the compositeness condition}
We attempt to remove the condition (\ref{composite condition1}) from the compositeness conditions of BHL. In this case the remaining condition (\ref{composite condition2}) becomes
\begin{equation}
\lambda(\Lambda)=0.\label{composite condition2-1}
\end{equation}
Even in the case, we can also path integrate out the Higgs field at $\Lambda$, and obtain an action having a non-local four-fermi interaction term,
\begin{equation}
I=\int d^4 x~{\cal L}_{qg}+S_{\mbox{{\scriptsize det}}}+S_{\mbox{{\scriptsize NL-4F}}},\label{non-local}
\end{equation}
where
\begin{eqnarray}
S_{\mbox{{\scriptsize det}}}&=&iTrLn\left\{K^{-1}(x-y)\right\},\label{det}\\
S_{\mbox{{\scriptsize NL-4F}}}&=&\int{\overline {\bf \Psi}}_{Li}(x){\bf Y}_{i}{\bf \Psi}_{Ri}(x)K(x-y){\overline {\bf \Psi}}_{Rj}(y){\bf Y}_{j}{\bf \Psi}_{Lj}(y),\label{non-local term}
\end{eqnarray}
and the repeated indices are summed and $\int$ means the integration with respect to the repeated arguments. We have defined:
\begin{eqnarray}
{\bf \Psi}_{Li}&=&\left(Q_{i}^{d},~~\widetilde{Q}_{i}^{u},~~L_{i}~\right)^T,~~\widetilde{Q}_{i}^{u}=i\sigma_{2} Q_{i}^{u\ast},\label{L}\\
{\bf \Psi}_{Ri}&=&\left(q_{i}^{d},~~q_{i}^{u\ast},~~l_{i}~\right)^{T},\label{R}\\
{\bf Y}_{i}&=&\mbox{diag}\left(y_{i}^{d},~y_{i}^{u},~y_{i}^{l}\right),
\end{eqnarray}
and
\begin{equation}
K^{-1}(x-y)=\left\{\partial^{2}_{x}+m^{2}-V(x)\right\}\delta^{4}(x-y),
\end{equation}
where
\begin{eqnarray}
V(x)&=&i\left(\partial^{\mu}A_{\mu}\right)+2iA_{\mu}\partial^{\mu}+A_{\mu}A^{\mu},\\
A_{\mu}&=&\frac{g'}{2}B_{\mu}+g_{2}\frac{\sigma_{a}}{2}W_{\mu}^{a}
\end{eqnarray}
and $B_{\mu}$ and $W_{\mu}^{a}$ are the $U(1)$ and $SU(2)_{L}$ gauge fields, respectively. $K(x-y)$ can be expressed as
\begin{eqnarray}
K(x-y)&=&iD(x-y)+\int iD(x-z)V(z)iD(z-y)\nonumber \\
&&+\int iD(x-z_1)V(z_1)iD(z_1-z_2)V(z_2)iD(z_2-y)+\cdots~~~~~
\end{eqnarray}
when the series converges, where
\begin{equation}
iD(x-y)=\int \frac{d^{4}k}{(2\pi)^{4}}\frac{e^{-ik(x-y)}}{m^{2}-k^{2}-i\epsilon}.
\end{equation}
When we give Eq.~(\ref{non-local}), we have also taken account of the Yukawa coupling constants for the other fermions than the top quark.

Therefore, we can describe the Higgs particle by using alternative actions at $\Lambda$ and below, one is the Higgs-Yukawa action of the standard model, and another is the action for the fermions that has the non-local four-fermi interaction term with other higher dimensional terms generated through the renormalization transformation from $\Lambda$. We note that the equation of motion for the Higgs field at $\Lambda$ yields
\begin{equation}
\Phi(x)=-K(x-y){\overline {\bf \Psi}}_{Ri}(y){\bf Y}_{i}{\bf \Psi}_{Li}(y)\label{composite field}
\end{equation}
in our case.

\begin{figure}[t]
 \leavevmode
 \epsfxsize= 100mm
 \epsfysize= 70mm
 \centerline{\epsfbox{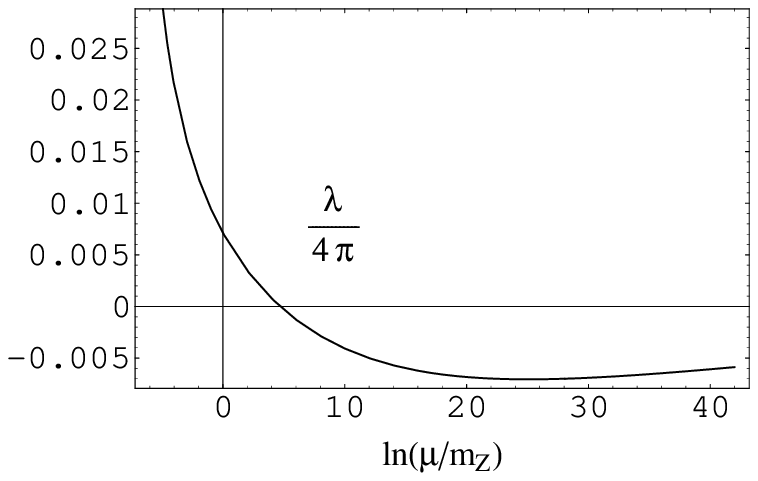}} 
 \caption{The RG evolution of $\alpha_{\lambda}=\lambda/(4\pi)$ for $\alpha_{\lambda}(\mu=10^{4}\mbox{ GeV})=0$ and $m_{t}=173.8$ GeV (${\overline m_{t}}=163.9$ GeV).}
 \label{ah}
\end{figure}

Now, what does occur above $\Lambda$? In the last section we saw that the condition (\ref{composite condition1}) means the breaking down of the Higgs-Yukawa system at $\Lambda$ due to the divergence of $y_{3}^{u}$. In our case the system does not break down at the scale in this sense because we have removed the condition (\ref{composite condition1}), however, the system is breaking down above $\Lambda$ from the viewpoint of the vacuum stability. Let us determine the running of $\lambda$. We can solve
\begin{equation}
\frac{d\alpha_{\lambda}}{dt}=\beta_{\alpha_{\lambda}}^{\mbox{{\scriptsize MS}}},~~\alpha_{\lambda}=\frac{\lambda}{4\pi}\label{lambda}
\end{equation}
using Eq.~(\ref{composite condition2-1}) as a boundary condition, where
\begin{equation}
4\pi \beta_{\alpha_{\lambda}}^{\mbox{{\scriptsize MS}}}=24\alpha_{\lambda}^{2}+ 12\alpha_{\lambda}\alpha_{t}-\frac{9}{5}\alpha_{\lambda}\alpha_{1}-9\alpha_{\lambda}\alpha_{2}+\frac{27}{200}\alpha_{1}^{2}+\frac{9}{20}\alpha_{1}\alpha_{2}+\frac{9}{8}\alpha_{2}^{2}-6\alpha_{t}^{2}\label{beta-a-lambda}
\end{equation}
is the 1-loop $\beta$ function for $\alpha_{\lambda}$ of the ${\overline {\rm MS}}$ scheme; we will confirm the validity of the use in Appendix A. As an example, we give the running of $\alpha_{\lambda}$ for $\Lambda=10^{4}$ GeV and $m_{t}=173.8$ GeV (${\overline m_{t}}=163.9$ GeV) in Fig.~\ref{ah}.
From the figure we can see that $\lambda$ becomes negative above $\Lambda$, and the Higgs potential becomes unbounded from below. This means that the states corresponding to the local minimum of the potential are unstable and the true vacuum state of the Higgs-Yukawa system cannot be defined.

After all, when we remove the condition (\ref{composite condition1}) from BHL's compositeness conditions, the two phenomena occur:

\begin{enumerate}
\renewcommand{\labelenumi}{(\Roman{enumi})}
\item the Higgs-Yukawa system of the standard model becomes equivalent to the non-local four-fermi interaction system at $\Lambda$.

\item the Higgs-Yukawa sector of the standard model becomes useless above $\Lambda$ because the vacuum state cannot be defined above the scale.
\end{enumerate}

\noindent
We consider that these indicate the compositeness of the Higgs particle.

Here, we give two comments. (1) In BHL's scenario, the 1-loop $\beta$ functions obtained in the perturbation theory did not permit an extrapolation all the way to $\Lambda$ because $\alpha_{t}$ diverges at $\Lambda$. In contrast to this, the perturbative expansion with respect to coupling constants is reliable near $\Lambda$ in our scenario because all the coupling constants are within the perturbative region near $\Lambda$. (2) We have discussed the vacuum stability based on the "tree potential" of the Higgs field assuming that higher dimensional terms in the effective potential of the Higgs field are negligible compared to the "tree potential" for $\phi^2 \to \infty$, where $\phi$ is the physical Higgs field. The verification of this assumption is not so easy in Wilson's renormalization scheme. We give a desired form for the effective lagrangian to make the assumption valid in Appendix B.

The removed condition (\ref{composite condition1}) was derived from the condition (\ref{condition for Z}) that coincides with another version of the compositeness condition discussed in many contexts~\cite{condition for Z}. If one identifies $Z(\mu)$ in the condition (\ref{condition for Z}) with the wave function renormalization constant of the Higgs field, the condition also implies the compositeness of the Higgs particle from the viewpoint of the following: the 1-particle pole part in the 2-point function of the Higgs field vanishes at $\Lambda$. In our scenario, the pole will not vanish at $\Lambda$. Then does our scenario have some significance for the compositeness of the Higgs particle bearing comparison with that the condition (\ref{condition for Z}) possesses? Let us consider this point from now.

One may consider that four-fermi interactions are low-energy effective interactions of some fundamental interaction. For the top-condensation models, this idea is studied in the so-called topcolor model~\cite{topcolor}, in which the local four-fermi interaction is rewritten into a form of the effective current-current interaction through Fierz transformation. Let us rewrite our non-local four-fermi interaction into such a form:
\begin{equation}
S_{\mbox{{\scriptsize NL-4F}}}=-\int J_{\mu ij}^{Lab}(x,y)J_{ij}^{R\mu ab}(x,y)\label{current},
\end{equation}
where
\begin{eqnarray}
J_{\mu ij}^{Lab}(x,y)&=&\frac{1}{\sqrt{2}}{\check y}_{i}^{a}{\check y}_{j}^{b}{\overline {\bf \Psi}}_{Li}^{a}(x)K(x-y)\gamma_{\mu}{\bf \Psi}_{Lj}^{b}(y),\\
J_{\mu ij}^{Rab}(x,y)&=&\frac{1}{\sqrt{2}}{\check y}_{j}^{b}{\check y}_{i}^{a}{\overline {\bf \Psi}}_{Rj}^{b}(y)\gamma^{\mu}{\bf \Psi}_{Ri}^{a}(x)
\end{eqnarray}
and the Latin indices $a\sim d$ distinguish the components of the right-hand sides of Eqs.~(\ref{L}) and (\ref{R}) including the difference of the colors of the quarks, ${\check y}_{i}^{a}$'s are defined by $({\check y}_{i}^{a})^{2}=y_{i}^{a}$ with ${\bf Y}_{i}^{ab}=y_{i}^{a}\delta^{ab}$. This current-current like interaction will be effectively produced by exchanges of heavy vector bosons represented by bi-local fields. First, we consider
\begin{equation}
S_{\mbox{{\scriptsize BV-I0}}}=\int A_{ij}^{\mu ab}(x,y)\left\{M^{2}J_{\mu ij}^{Lab}(x,y)-J_{\mu ij}^{Rab}(x,y)\right\},\label{new interaction}
\end{equation}
where $A^{\mu ab}_{ij}(x,y)=A_{ji}^{\mu ba\ast}(y,x)$'s are the bi-local vector fields that are $SU(2)_{L}$ singlets, $M^{2}$ is a parameter having mass dimension 2. The interaction Eq.~(\ref{new interaction}) is invariant under $SU(3)_{c}\times SU(2)_{L}\times U(1)_{Y}$ gauge transformations if the bi-local vector fields transform as
\begin{equation}
A_{ij}^{\mu}(x,y)\to U_{R}(x)U_{g}(x)A_{ij}^{\mu}(x,y)U_{g}^{\dag}(y)U_{R}^{\ast}(y),\label{adjoint}
\end{equation}
where $U_{R}(x)$ and $U_{g}(x)$ are the matrices of the $U(1)$ and the $SU(3)_{c}$ gauge transformations acting on ${\bf \Psi}_{R}(x)$, respectively. We introduce the Fourier transformation for the bi-local vector fields:
\begin{equation}
A_{ij}^{\mu}(x,y)=\frac{1}{(\sqrt{2\pi})^{4}}\int d^{4}q~e^{-iqr}A_{ij}^{\mu}(X;q),\label{infinite number of fields}
\end{equation}
where $r=x-y$ and $X=(x+y)/2$, and so the bi-local fields are expressed by an infinite number of the local fields labeld by $q$. We suppose that
\begin{equation}
~_{I}\langle0|\,T\,A_{ij}^{\mu ab}(X_{1};q_{1})A_{kl}^{\nu cd\ast}(X_{2};q_{2})|0\rangle_{I}=\delta(q_{1}-q_{2})g^{\mu\nu}\delta_{ik}^{ac}\delta_{jl}^{bd}\int \frac{d^{4}P}{i(2\pi)^{4}}\frac{e^{-iP(X_{1}-X_{2})}}{M^{2}-P^{2}-i\epsilon}
\end{equation}
in the interaction picture, where $\delta^{ac}_{ik}$ and $\delta(x)$ represent $\delta^{ac}\delta_{ik}$ and $\delta^{4}(x)$, respectively. Then, we have
\begin{equation}
~_{I}\langle0|\,T\,A_{ij}^{\mu ab}(x_{1},y_{1})A_{kl}^{\nu cd\ast}(x_{2},y_{2})\,|0\rangle_{I} \to \frac{1}{iM^{2}}\delta(x_{1}-x_{2})\delta(y_{1}-y_{2})g^{\mu\nu}\delta_{ik}^{ac}\delta_{jl}^{bd}
\end{equation}
for $P^{2}\ll M^{2}$, and the interaction $S_{\mbox{{\scriptsize BV-I0}}}$ will induce $S_{\mbox{{\scriptsize NL-4F}}}$ of Eq.~(\ref{current}) in this case.

We note that the interaction $S_{\mbox{{\scriptsize BV-I0}}}$ will also induce other four-fermi interactions than $S_{\mbox{{\scriptsize NL-4F}}}$, and also we cannot rely the perturbative expansions with respect to ${\check y}_{i}^{a}{\check y}_{j}^{b}M^{2}$'s. We next give an example in which bi-local vector fields can be pass integrated out exactly~\cite{bilocal}, and the four-fermi interaction obtained through the integration is only $S_{\mbox{{\scriptsize NL-4F}}}$. We introduce one more kind of the bi-local vector fields $C^{\mu ab}_{ij}(x,y)=C^{\mu ba\ast}_{ji}(y,x)$, and consider
\begin{eqnarray}
{\widehat S}_{\mbox{{\scriptsize BV}}}&=&\int tr\left[\left\{D_{\mu}A_{\nu}(x,y)\right\}^{\dag}D^{\mu}A^{\nu}(x,y)-M^{2}A^{\dag}_{\mu}(x,y)A^{\mu}(x,y)\right]\nonumber \\
&&-\int tr\Big[A_{\mu}(x,y)\to C_{\mu}(x,y)\Big]+S_{\mbox{{\scriptsize BV-I}}},\label{S_BV}
\end{eqnarray}
where
\begin{eqnarray}
S_{\mbox{{\scriptsize BV-I}}}&=&\int\Big[A_{ij}^{\mu ab}(x,y)\left\{M^{2}J_{\mu ij}^{Lab}(x,y)-J_{\mu ij}^{Rab}(x,y)\right\}\nonumber \\
&&-C_{ij}^{\mu ab}(x,y)\left\{M^{2}J_{\mu ij}^{Lab}(x,y)+J_{\mu ij}^{Rab}(x,y)\right\}\Big],
\end{eqnarray}
\begin{equation}
\left\{D^{\mu}A^{\nu}(x,y)\right\}^{ab}_{ij}=\left\{\delta^{ac}_{ik}\delta^{bd}_{jl}\partial_{X}^{\mu}-iB^{\mu ab:cd}_{ij:kl}(x,y)\right\}A_{kl}^{\nu cd}(x,y)\label{covariant derivative for H}
\end{equation}
with
\begin{equation}
\partial_{X}^{\mu}=\partial_{x}^{\mu}+\partial_{y}^{\mu},~~~B_{~\;ij:kl}^{\mu ab:cd}(x,y)=B_{ik}^{\mu ac}(x)\delta^{bd}_{jl}-\delta^{ac}_{ik}B^{\mu bd}_{jl}(y).
\end{equation}
$B_{ik}^{\mu ac}(x)=B_{ki}^{\mu ca \ast}(x)$'s are components of a matrix composed of $U(1)_{Y}$ and $SU(3)_{c}$ gauge fields multiplied by ${g'}$ and $g_{3}$, respectively, which make Eq.~(\ref{covariant derivative for H}) to be gauge covariant under Eq.~(\ref{adjoint}). The definition for $D^{\mu}C^{\nu}(x,y)$ is just the same with one for $D^{\mu}A^{\nu}(x,y)$. In Eq.~(\ref{S_BV}), $tr$ means the usual trace for the matrix having two Latin indices\footnote{We will also define the trace for matrices that have four Latin indices later.}, and we have written $\{D^{\mu}A^{\nu}(x,y)\}^{\dag ab}_{ij}$ as $\{D^{\mu}A^{\nu}(x,y)\}^{ba\ast}_{ji}$. We can integrate out the bi-local vector fields exactly because ${\widehat S}_{\mbox{{\scriptsize BV}}}$ is quadratic with respect to the fields. Let us rewrite ${\widehat S}_{\mbox{{\scriptsize BV}}}$ into
\begin{eqnarray}
\lefteqn{{\widehat S}_{\mbox{{\scriptsize BV}}}}\nonumber \\
&=&-\int\Bigl[\left\{A_{\mu ij}^{ab\ast}(x_{1},y_{1})-\xi_{\mu mn}^{ef}(x_{3},y_{3})\Delta^{ef:ab}_{mn:ij}(x_{3},y_{3}:x_{1},y_{1})\right\}\nonumber \\
&&\cdot~\Delta^{-1~ab:cd}_{ij:kl}(x_{1},y_{1}:x_{2},y_{2})\left\{A^{\mu cd}_{kl}(x_{2},y_{2})-\Delta^{cd:gh}_{kl:op}(x_{2},y_{2}:x_{4},y_{4})\xi^{\mu gh\ast}_{op}(x_{4},y_{4})\right\}\nonumber \\
&&-\left(A_{\mu}\to C_{\mu},~~\xi_{\mu}\to \eta_{\mu}\right)\Bigr]+S_{\mbox{{\scriptsize B-4F}}},\label{reduced}
\end{eqnarray}
where
\begin{equation}
S_{\mbox{{\scriptsize B-4F}}}=-\int J_{\mu ij}^{Lab}(x_{1},y_{1})M^{2}{\widetilde \Delta}^{ab:cd}_{ij:kl}(x_{1},y_{1}:x_{2},y_{2})J_{kl}^{R\mu cd}(x_{2},y_{2}),\label{reduced2}
\end{equation}
\begin{eqnarray}
\xi_{\mu ij}^{ab}(x,y)&=&\frac{1}{2}\left\{M^{2}J_{\mu ij}^{Lab}(x,y)-J_{\mu ij}^{Rab}(x,y)\right\},\\
\eta_{\mu ij}^{ab}(x,y)&=&\frac{1}{2}\left\{M^{2}J_{\mu ij}^{Lab}(x,y)+J_{\mu ij}^{Rab}(x,y)\right\},~~~~~
\end{eqnarray}
and
\begin{equation}
\Delta^{-1~ab:cd}_{ij:kl}(x_{1},y_{1}:x_{2},y_{2})=iD^{-1~ab:cd}_{ij:kl}(x_{1},y_{1}:x_{2},y_{2})-V^{ab:cd}_{ij:kl}(x_{1},y_{1}:x_{2},y_{2})
\end{equation}
with 
\begin{equation}
iD^{-1~ab:cd}_{ij:kl}(x_{1},y_{1}:x_{2},y_{2})=\delta^{ac}_{ik}\delta^{bd}_{jl}(\partial_{X_{1}}^{2}+M^{2})\delta(x_{1}-x_{2})\delta(y_{1}-y_{2})
\end{equation}
and
\begin{eqnarray}
\lefteqn{V^{ab:cd}_{ij:kl}(x_{1},y_{1}:x_{2},y_{2})}\nonumber \\
&=&\left[i\{\partial^{\mu}_{X_{1}}B^{ab:cd}_{\mu ij:kl}(x_{1},y_{1})\}+2iB^{ab:cd}_{\mu ij:kl}(x_{1},y_{1})\partial^{\mu}_{X_{1}}+B^{ab:ef}_{\mu ij:mn}(x_{1},y_{1})B^{\mu ef:cd}_{mn:kl}(x_{1},y_{1})\right]\nonumber \\
&&\cdot~\delta(x_{1}-x_{2})\delta(y_{1}-y_{2}).~~~~~~~~
\end{eqnarray}
The inverse of $\Delta^{-1 ab:cd}_{ij:kl}(x_{1},y_{1}:x_{2},y_{2})$ is defined by
\begin{eqnarray}
&&\int\Delta^{ab:ef}_{ij:mn}(x_{1},y_{1}:x_{3},y_{3})\Delta^{-1~ef:cd}_{mn:kl}(x_{3},y_{3}:x_{2},y_{2})\nonumber \\
&&=\int\Delta^{-1~ab:ef}_{ij:mn}(x_{1},y_{1}:x_{3},y_{3})\Delta^{ef:cd}_{mn:kl}(x_{3},y_{3}:x_{2},y_{2})\nonumber \\
&&=\delta^{ac}_{ik}\delta^{bd}_{jl}\delta(x_{1}-x_{2})\delta(y_{1}-y_{2}).
\end{eqnarray}
We can express it as
\begin{equation}
\Delta=iD+iDViD+iDViDViD+\cdots
\end{equation}
when the series converges, where
\begin{equation}
iD^{ab:cd}_{ij:kl}(x_{1},y_{1}:x_{2},y_{2})=\delta^{ac}_{ik}\delta^{bd}_{jl}\delta(r_{1}-r_{2})\int \frac{d^{4}P}{(2\pi)^{4}}\frac{e^{-iP(X_{1}-X_{2})}}{M^{2}-P^{2}-i\epsilon}
\end{equation}
and we have omitted the indices with defining the product of the matrices $M$ and $N$ having four indices as
\begin{equation}
(MN)^{ab:cd}_{ij:kl}(x_{1},y_{1}:x_{2},y_{2})=\int M^{ab:ef}_{ij:mn}(x_{1},y_{1}:x_{3},y_{3})N^{ef:cd}_{mn:kl}(x_{3},y_{3}:x_{2},y_{2}).\label{product}
\end{equation}
It can be shown that
\begin{equation}
\Delta^{ab:cd}_{ij:kl}(x_{1},y_{1}:x_{2},y_{2})=\Delta^{cd:ab\ast}_{kl:ij}(x_{2},y_{2}:x_{1},y_{1}).
\end{equation}
We have also defined ${\widetilde \Delta}^{ab:cd}_{ij:kl}(x_{1},y_{1}:x_{2},y_{2})$ as
\begin{equation}
{\widetilde \Delta}^{ab:cd}_{ij:kl}(x_{1},y_{1}:x_{2},y_{2})=\frac{1}{2}\left\{\Delta^{ab:cd}_{ij:kl}(x_{1},y_{1}:x_{2},y_{2})+\Delta^{dc:ba}_{lk:ji}(y_{2},x_{2}:y_{1},x_{1})\right\},
\end{equation}
which satisfies
\begin{equation}
{\widetilde \Delta}^{ab:cd}_{ij:kl}(x_{1},y_{1}:x_{2},y_{2})={\widetilde \Delta}^{dc:ba}_{lk:ji}(y_{2},x_{2}:y_{1},x_{1})={\widetilde \Delta}^{ba:dc\ast}_{ji:lk}(y_{1},x_{1}:y_{2},x_{2}).
\end{equation}
From Eq.~(\ref{reduced}), we have
\begin{equation}
\frac{1}{i}Ln\int {\cal D}A{\cal D}A^{\ast}{\cal D}C{\cal D}C^{\ast}\exp{\left[iS_{\mbox{{\scriptsize BV}}}\right]}=S_{\mbox{{\scriptsize B-4F}}},
\end{equation}
where
\begin{equation}
S_{\mbox{{\scriptsize BV}}}={\widehat S}_{\mbox{{\scriptsize BV}}}-2iTrLn\left[\Delta^{-1 ab:cd}_{ij:kl}(x_{1},y_{1}:x_{2},y_{2})\right]\label{S_BV-new}
\end{equation}
and $Ln[\Delta^{-1}]$ is defined by using the power series of the matrices having four indices with Eq.~(\ref{product}), and we have written as $Tr[N^{ab:cd}_{ij:kl}(x_{1},y_{1}:x_{2},y_{2})]=\int N^{ab:ab}_{ij:ij}(x_{1},y_{1}:x_{1},y_{1})$. Note that
\begin{equation}
M^{2}{\widetilde \Delta}^{ab:cd}_{ij:kl}(x_{1},y_{1}:x_{2},y_{2})\to \delta^{ac}_{ik}\delta^{bd}_{jl}\delta(x_{1}-x_{2})\delta(y_{1}-y_{2})
\end{equation}
for $P^{2} \ll M^{2} \to \infty$, and so $S_{\mbox{{\scriptsize B-4F}}}$ becomes $S_{\mbox{{\scriptsize NL-4F}}}$ of Eq.~(\ref{current}) in this case. Thus, we can consider that our non-local four-fermi interaction Eq.~(\ref{non-local term}) is the low-energy effective interaction induced in the system of $S_{\mbox{{\scriptsize BV}}}$ by assuming that $M^{2}$ becomes large as the energy scale approaches to $\Lambda$.

On the other hand, we can rewrite $S_{\mbox{{\scriptsize B-4F}}}$ of Eq.~(\ref{reduced2}) into
\begin{eqnarray}
S_{\mbox{{\scriptsize B-4F}}}&=&\int{\overline {\bf \Psi}}_{Li}^{a}(x_{1}){\bf \Psi}_{Rk}^{c}(x_{2}){\check y}_{i}^{a}{\check y}_{j}^{b}{\check y}_{k}^{c}{\check y}_{l}^{d}K(x_{1}-y_{1})M^{2}{\widetilde \Delta}^{ab:cd}_{ij:kl}(x_{1},y_{1}:x_{2},y_{2})\nonumber \\
&&\cdot{\overline {\bf \Psi}}_{Rl}^{d}(y_{2}){\bf \Psi}_{Lj}^{b}(y_{1})\label{another form of B-4F}
\end{eqnarray}
through Fierz transformation, which becomes the original expression of $S{\mbox{{\scriptsize NL-4F}}}$ of Eq.~(\ref{non-local term}) for $P^{2} \ll M^{2} \to \infty$. The form of Eq.~(\ref{another form of B-4F}) suggests that $S{\mbox{{\scriptsize BV}}}$ is equivalent to some action written in terms of scalar fields. We note
\begin{eqnarray}
&&\frac{1}{i}Ln\int {\cal D}\chi{\cal D}\chi^{\dag}{\rm exp}\left[
-i\int\left\{\chi_{ij}^{ab\dag}(x_{1},x_{2})+{\overline {\bf \Psi}}_{Li}^{a}(x_{3}){\check y}_{i}^{a}{\check y}_{j}^{b}{\bf \Psi}_{Rj}^{b}(x_{2})K(x_{3}-x_{1})\right\}\right.\nonumber \\
&&\cdot~K^{-1~ab:cd}_{ij:kl}(x_{1},x_{2}:y_{1},y_{2})\left\{\chi_{kl}^{cd}(y_{1},y_{2})+K(y_{1}-y_{3}){\overline {\bf \Psi}}_{Rl}^{d}(y_{2}){\check y}_{l}^{d}{\check y}_{k}^{c}{\bf \Psi}_{Lk}^{c}(y_{3})\right\}\biggr]\nonumber \\
&&=iTrLn\left[K^{-1~ab:cd}_{ij:kl}(x_{1},x_{2}:y_{1},y_{2})\right],\label{path integral}
\end{eqnarray}
where $\chi_{ij}^{ab}(x_{1},x_{2})=\chi_{ji}^{ba\ast}(x_{2},x_{1})$'s are the bi-local scalar doublets, and
\begin{eqnarray}
\lefteqn{K^{-1~ab:cd}_{ij:kl}(x_{1},x_{2}:y_{1},y_{2})}\nonumber \\
&&=\int K^{-1}(x_{1}-x_{4})K(x_{4}-y_{4})K^{-1}(y_{4}-y_{1})M^{2}{\widetilde \Delta}^{ac:bd}_{ik:jl}(x_{4},y_{4}:x_{2},y_{2}),~~~~~
\end{eqnarray}
which satisfies $K^{-1~ab:cd~\dag}_{ij:kl}(x_{1},x_{2}:y_{1},y_{2})=K^{-1~cd:ab}_{kl:ij}(y_{1},y_{2}:x_{1},x_{2})$. Because the integrand of the path integral of the left-hand sides of Eq.~(\ref{path integral}) is equal to
\begin{equation}
\exp{\left[i\left(S_{\mbox{{\scriptsize $B\chi$}}}^{\mbox{{\scriptsize kin.}}}+S_{\mbox{{\scriptsize $B\chi$}}}^{\mbox{{\scriptsize Y}}}-S_{\mbox{{\scriptsize B-4F}}}\right)\right]},
\end{equation}
where
\begin{eqnarray}
S_{\mbox{{\scriptsize $B\chi$}}}^{\mbox{{\scriptsize kin.}}}&=&-\int\chi^{ab\dag}_{ij}(x_{1},x_{2})K^{-1~ab:cd}_{ij:kl}(x_{1},x_{2}:y_{1},y_{2})\chi_{kl}^{cd}(y_{1},y_{2}),\\
S_{\mbox{{\scriptsize $B\chi$}}}^{\mbox{{\scriptsize Y}}}&=&-\int\left[{\overline {\bf \Psi}}_{Li}^{a}(x_{3}){\check y}_{i}^{a}{\check y}_{j}^{b}{\bf \Psi}_{Rj}^{b}(x_{2})K(x_{3}-x_{1})K^{-1~ab:cd}_{ij:kl}(x_{1},x_{2}:y_{1},y_{2})\chi_{kl}^{cd}(y_{1},y_{2})\right.\nonumber \\
&&+\mbox{ h.c.}\Bigr],
\end{eqnarray}
we have
\begin{equation}
\frac{1}{i}Ln\int {\cal D}\chi{\cal D}\chi^{\dag}\exp{\left[iS_{\mbox{{\scriptsize $B\chi$}}}\right]}=S_{\mbox{{\scriptsize B-4F}}}=\frac{1}{i}Ln\int {\cal D}A{\cal D}A^{\ast}{\cal D}C{\cal D}C^{\ast}\exp{\left[iS_{\mbox{{\scriptsize BV}}}\right]},~
\end{equation}
where
\begin{equation}
S_{\mbox{{\scriptsize $B\chi$}}}=S_{\mbox{{\scriptsize $B\chi$}}}^{\mbox{{\scriptsize kin.}}}+S_{\mbox{{\scriptsize $B\chi$}}}^{\mbox{{\scriptsize Y}}}-iTrLn\left[K^{-1~ab:cd}_{ij:kl}(x_{1},x_{2}:y_{1},y_{2})\right].\label{chi}
\end{equation}
Thus, $S_{\mbox{{\scriptsize BV}}}$ is equivalent to the action for the bi-local scalar fields $S_{\mbox{{\scriptsize $B\chi$}}}$. It will be interest to see what $S_{\mbox{{\scriptsize $B\chi$}}}$ becomes for $P^{2} \ll M^{2} \to \infty$. When we set $\chi_{ii}^{aa}(x_{1},x_{2})$ to $\Phi(x_{1})$, i.e., to be independent of $x_{2}$, $S_{\mbox{{\scriptsize $B\chi$}}}^{\mbox{{\scriptsize kin.}}}$ and $S_{\mbox{{\scriptsize $B\chi$}}}^{\mbox{{\scriptsize Y}}}$ become the kinetic term and the Yukawa interaction terms of the Higgs-Yukawa sector of the standard model, respectively.

We have given an example of the action for the physics above $\Lambda$, $S_{\mbox{{\scriptsize BV}}}$ or $S_{\mbox{{\scriptsize $B\chi$}}}$. The actions are equivalent to the action for fermions having $S_{\mbox{{\scriptsize B-4F}}}$ that becomes the Higgs-Yukawa sector of the standard model at $\Lambda$ in the limit $P^{2} \ll M^{2} \to \infty$\footnote{To be exact, $S_{\mbox{{\scriptsize B-4F}}}+S_{\mbox{{\scriptsize det}}}$ becomes equivalent to the sector in this limit, where $S_{\mbox{{\scriptsize det}}}$ is of Eq.~(\ref{det}). So, the action above $\Lambda$ is $S_{\mbox{{\scriptsize $B\chi$}}}+S_{\mbox{{\scriptsize det}}}$ or $S_{\mbox{{\scriptsize BV}}}+S_{\mbox{{\scriptsize det}}}$.}. It is important to note that $S_{\mbox{{\scriptsize BV}}}$ or $S_{\mbox{{\scriptsize $B\chi$}}}$ is not exactly equivalent to the action for the fermions having $S_{\mbox{{\scriptsize NL-4F}}}$ but induces it effectively. If they are exactly equivalent, then we can describe the physics above $\Lambda$ using the Higgs-Yukawa sector of the standard model alternatively, because $S_{\mbox{{\scriptsize NL-4F}}} (+S_{\mbox{{\scriptsize det}}})$ is equivalent to the sector. Then the vacuum of the system will be ill defined above $\Lambda$.

It should be emphasized that $S_{\mbox{{\scriptsize BV}}}$ could not be rewritten into the action for the single elementary local scalar field but the bi-local scalar fields. In general, although we cannot rewrite an action into another one through the auxiliary field method, the two actions might be equivalent to each other in the sense that the vertex functions obtained from both actions are identical with each other\footnote{In Refs.~\cite{equivalence-1,equivalence-2}, the equivalence between some generalized NJL system and the unconstrained Higgs-Yukawa system is shown in this way in the large $N_{c}$ limit.}. However, there will exist no action for a single elementary local scalar field that is equivalent to $S_{\mbox{{\scriptsize BV}}}$, because it has been already shown that $S_{\mbox{{\scriptsize BV}}}$ is equivalent to action for the bi-local scalar fields that are represented by an infinite number of local fields like Eq.~(\ref{infinite number of fields}). Therefore, we can say that the scale $\Lambda$ is the critical scale above which the Higgs particle cannot be described by the action written in terms of the elementary local field for the Higgs particle but described by one for the constituent particles of the Higgs particle or some action for bi-local scalar fields. From this, we consider that the scale $\Lambda$ has a qualification of the compositeness scale.

\section{Higgs boson mass}
Now, we predict the Higgs boson mass in our scenario depending on $\Lambda$. The mass is obtained from
\begin{equation}
{\overline m_{H}}(\mu)=\sqrt{8\pi\alpha_{\lambda}(\mu)}\,v\label{Higgs mass}
\end{equation}
because $\alpha_{\lambda}(\mu)$ has been determined thanks to the condition (\ref{composite condition2-1}). The results are shown in Table I.
\begin{table}
Table I. The Higgs boson masses obtained from Eq.~(\ref{Higgs mass}). ${\overline m_{t}}$'s are defined by Eq.~(\ref{pole}). The values having the superscript $*$ are obtained using Eqs.~(\ref{bm}) and (\ref{bl}). The results for $\Lambda =10^{19}$ GeV are not reliable because our $\beta$ function is not reliable for $\Lambda \simeq m_{Pl}$ (See Appendix A). We must take notice that ${\overline m_{H}}({\overline m_{H}})$'s are not the pole masses.
\let\tabularsize

\begin{center}
\begin{tabular}{c|c|c|c}\hline \hline
$\Lambda$[GeV]&\begin{minipage}{3.5cm}
\begin{center}
${\overline m_{H}}({\overline m_{H}})$ [GeV] for\\
$m_{t}=168.6$ GeV\\
(${\overline m_{t}}=159.0$ GeV)
\end{center}
\end{minipage}&\begin{minipage}{3.5cm}
\begin{center}
${\overline m_{H}}({\overline m_{H}})$ [GeV] for\\
$m_{t}=173.8$ GeV\\
(${\overline m_{t}}=163.9$ GeV)
\end{center}
\end{minipage}&\begin{minipage}{3.5cm}
\begin{center}
${\overline m_{H}}({\overline m_{H}})$ [GeV] for\\
$m_{t}=179.0$ GeV\\
(${\overline m_{t}}=168.9$ GeV)
\end{center}
\end{minipage}\\
\hline
$(10^{19})$&(126)&(136)&(146)\\
$10^{15}$&126&135&145\\
$10^{10}$&122&131&140\\
$10^{4}$&$96$&$102$&$108$\\
$10^{3}$&$81^{*}$&$85^{*}$&$90^{*}$\\
\hline
\end{tabular}
\end{center}
\end{table}
For $\Lambda=10^{3}$ GeV, we have used the $\beta$ functions of our cutoff scheme that will be defined in Appendix A, namely, Eqs.~(\ref{bm}) and (\ref{bl})\footnote{Although we use $\beta_{\alpha_{\lambda}}^{\mbox{{\scriptsize MS}}}$ for $\Lambda=10^{3}$ GeV, the results does not differ from the values in Table I more than $0.1$ GeV.}. On the other hand, for $\Lambda=10^{19},~10^{15},~10^{10},~10^{4}$ GeV, we have used Eq.~(\ref{beta-a-lambda}) approximately. This is because a problem of fine-tuning arises for $\Lambda \gg m_{z}$ when we use the $\beta$ functions of our cutoff scheme (See Appendix A). We will show that our $\beta$ function for $\alpha_{\lambda}$ reduces to $\beta_{\alpha_{\lambda}}^{\mbox{{\scriptsize MS}}}$ under an approximation. We will also see that our $\beta$ functions are not reliable for $\Lambda \simeq m_{Pl}$, and so the results for $\Lambda=10^{19}$ GeV in Table I are not reliable .

We must take notice that ${\overline m_{H}}({\overline m_{H}})$'s are not the pole masses. The pole mass $m_{H}$ and ${\overline m_{H}}({\overline m_{H}})$ are related by
\begin{equation}
{\overline m^{2}_{H}}(\mu)=m^{2}_{H}(1+\delta_{H}(\mu)),\label{pole-H}
\end{equation}
where $\delta_{H}(\mu)$ depends on $m_{H}$~\cite{correction1,correction2}. From the results shown in Ref.~\cite{correction1}, we can see that $\delta_{H}(\mu)$ decreases with increasing $m_{H}$ for the values of ${\overline m_{H}}({\overline m_{H}})$ in Table I\footnote{We can see that $0.05<\delta_{H}(\mu)\lsim 1$.}, and so the pole masses will be smaller than ${\overline m_{H}}({\overline m_{H}})$.

To conclude this section, we give two comments. (1) Our way used to predict the Higgs boson mass is the same with one to yield the ordinary stability bound of the mass~\cite{stability bound-1,stability bound-2}, except that we treat our model as a cutoff model. In ordinary discussion of the stability bound, the lower limits on the mass are derived from the requirement that the Higgs potential is bounded below up to a certain scale where some new physics appears. The compositeness scale $\Lambda$ of our scenario corresponds to such a scale. The value of $\Lambda$ is a free parameter of our model, however, we consider that $\Lambda$ is not so large compared to $m_{Z}$ because if $\Lambda \gg m_{Z}$, we encounter a problem of the fine-tuning as we will detail in Appendix A. This problem can be considered a kind of the problem of hierarchy. The absence of such a fine-tuning may require $\Lambda \lsim O(10^{3}-10^{4} \mbox{GeV})$~\cite{stability bound-3}. (2) The Higgs boson mass is sensitive to the variation of $m_{t}$ relative to one of $\Lambda$. This seems to support the idea that the Higgs particle is composed of the top quarks.

\section{Summary and discussions}
The compositeness conditions of BHL, Eqs.~(\ref{composite condition1}) and (\ref{composite condition2}) lead to the two phenomena (i) and (ii) in \S 3. These are considered to indicate the compositeness of the Higgs particle. Now that the experimental value of the top quark mass contradicts the condition (\ref{composite condition1}), we must give up the scenario of BHL without some improvement. In this paper we considered what occurs when we remove Eq.~(\ref{composite condition1}).

If one removes Eq.~(\ref{composite condition1}) from BHL's compositeness conditions, the remaining condition (\ref{composite condition2}) becomes Eq.~(\ref{composite condition2-1}). This condition leads to the two phenomena (I) and (II) in \S 3. We consider that these, together with each other, also indicate the compositeness of the Higgs particle.

The phenomenon (I) may not be so surprising one. In Refs.~\cite{equivalence-1,equivalence-2}, it is shown in the large $N_{c}$ limit that even in the absence of any conditions for $\lambda(\mu)$ there exists the generalized Nambu-Jona-Lasinio (NJL) system to be equivalent to the Higgs-Yukawa system in the sense that the vertex functions obtained in both systems are identical with each other. If the equivalence is exact, we should say that Eq.~(\ref{composite condition2-1}) is the condition to guarantee the statement (II), and then the condition also determine the type of the generalized NJL lagrangian that is equivalent to the Higgs-Yukawa lagrangian.

The condition (\ref{condition for Z}), which arises the removed condition (\ref{composite condition1}), can be considered the compositeness condition also from the viewpoint of that the 1-particle pole part in the 2-point function of the Higgs field vanishes at $\Lambda$. In our scenario, the pole will not vanish at $\Lambda$, and so we cannot say $\Lambda$ to be the scale at which the asymptotic field for the Higgs particle $\sqrt{Z}\phi_{as}$ vanishes. However, the scale $\Lambda$ is considered to be the compositeness scale of the Higgs particle from another point of view as follows. In the ordinary discussion of the vacuum stability, it would be considered that some new physics appears before the Higgs potential becomes unbounded below, and the Higgs-Yukawa system supplemented with some new interaction does not break down. In our scenario we consider that the Higgs-Yukawa system breaks down above $\Lambda$, and the physics above the scale is described by the action that does not have the Higgs-Yukawa sector of the standard model. This is naturally understood when we consider the non-local four-fermi interaction Eq.~(\ref{non-local term}) to be a low-energy (of course higher than $\Lambda$) effective interaction produced by exchanges of heavy vector bosons, which are represented by the bi-local fields in our scenario because of the non-locality of the four-fermi interaction. Then even if we rewrite the action for the vector fields into an action for the scalar fields, the action will not be for a single local scalar field but for the bi-local scalar fields as discussed in \S 3. Therefore, we can say that the scale $\Lambda$ is the critical scale above which the Higgs particle cannot be described by the action written in terms of the local elementary field for Higgs particle but described by one for the constituent particles of the Higgs particle or some action for bi-local scalar fields. Contrasting with this, at $\Lambda$ and below, the Higgs particle can be described by alternative actions, one is the action for the constituent particles of the Higgs particle, namely, the action having the non-local four-fermi term Eq.~(\ref{non-local term}) with higher dimensional terms generated through the RG transformation from $\Lambda$, and another is the Higgs-Yukawa action, which is of course written in terms of the elementary local field for the Higgs particle. From this point of view, we consider that the scale $\Lambda$ has a qualification for the compositeness scale.

Now, we note that if the equivalence between some generalized NJL system and the unconstrained Higgs-Yukawa system discussed in Refs.~\cite{equivalence-1,equivalence-2} is exact, we can say as follows in our criterion for the compositeness: the Higgs particle described by the Higgs-Yukawa sector is the composite particle whenever the sector breaks down at a high-energy from any reason. So, it would appear that the Higgs particle is already fated to be the composite particle when we adopt the Higgs-Yukawa system to describe the low-energy physics of the Higgs particle or the symmetry breaking of the gauge theory.

It remains to be shown whether the vacuum of the system of $S_{\mbox{{\scriptsize BV}}}$ or $S_{\mbox{{\scriptsize $B\chi$}}}$ given in \S 3 is well defined or not. We will not address to this question here~\cite{stability of bi-local system}. Our example of the new theory above $\Lambda$ may be cheap a little to consider it a realistic model; it is sufficient to explain our criterion for the compositeness of the Higgs particle, though. We may be able to construct the new theory above $\Lambda$ based on some gauge symmetry that requires the bi-local gauge fields~\cite{bi-local gauge fields}. Then, the following should be required in general. (1) the vacuum is well defined above $\Lambda$. (2) the masses of the bi-local vector fields become large as the energy scale approaches to $\Lambda$. (3) the new theory breaks down at $\Lambda$ and below. Detailed studies of the theory above $\Lambda$ remains as future works.

\section*{Acknowledgements}
The author expresses his sincere thanks to Prof.~S.~Naka for useful comments and discussions. He is grateful to other members of his laboratory for their comments and  encouragement. He thanks Prof.~H.~Nakano for pointing out Ref.~\cite{equivalence-1}. He also thanks Dr.~C.~Itoi for discussions on the vacuum stability in Wilson's renormalization approach.

\vspace{0.2cm}

\appendix
\vspace{0.5cm}
\noindent
{\Large \bf Appendix}
\section{Renormalization transformation and $\beta$-functions}
In this Appendix, we define the renormalization transformation for the action and give the $\beta$ functions of our model with having Wilson's renormalization approach in mind. Our definition is about the same with one of Ref.~\cite{BKMN}. In Ref.~\cite{BKMN}, the $1/N_{c}$ expansion was used, and then one can elude the problem of the gauge invariance in the cutoff theory accidentally. We will use the perturbative expansion with respect to coupling constants; we can use the perturbation in our scenario as we have mentioned in \S 3. In our case the problem of the gauge invariance will be serious and remains unsolved in this paper. For this reason we give the $\beta$ functions only for $\lambda$ and $m^{2}$, and assume that for the other coupling constants, the $\beta$ functions of the ${\overline {\rm MS}}$ scheme in the continuous theory can be used approximately.

Our definition of the renormalization transformation for the action, $S \to S+\Delta S$, is as follows.

Step I: We define the transformation coming from the path integration over the fields with the Euclidean momentum in the infinitesimal interval $\mu^{2}\ge k^{2} \ge \mu^{2}-\Delta\mu^{2}$:
\begin{eqnarray}
\lefteqn{S[b_{0},f_{0},{\overline f_{0}};\mu]+{\widetilde \Delta} S[b_{0},f_{0},{\overline f_{0}};\mu]}\nonumber \\
&=&\frac{1}{i}\int^{\mu^{2}\sim \mu^{2}-\Delta\mu^{2}}{\cal D}[b]{\cal D}[f]{\cal D}[{\overline f}]\exp\left\{i\int d^{4}x {\cal L}(b_{0}+b, f_{0}+f, {\overline f_{0}}+{\overline f};\mu)\right\},~~~~~~\label{step-1}
\end{eqnarray}
where ${\cal L}(\mu)$ is the effective lagrangian at a scale $\mu$, $[b]$ and $[f]$ denote all fields for bosons and fermions contained in ${\cal L}(\mu)$, respectively, and $b_{0}$, $f_{0}$ and ${\overline f_{0}}$ are the fields that obey the equations of motion obtained from ${\cal L}(\mu)$. In the 1-loop approximation, we have
\begin{equation}
{\widetilde \Delta} S=\frac{i}{2}{\widetilde {\rm sTr}}\,{\rm Ln}\left[\frac{1}{2} F.T.\!\left\{\left.\frac{\delta^{2}S[b,f,{\overline f};\mu]}{\delta \varphi_{\alpha}(y) \delta \varphi_{\beta}(z)}~\right|_{0}\right\}\right],\label{1-loop}
\end{equation}
where $\varphi_{\alpha}$'s denote the field contained in ${\cal L}(\mu)$ and the Greek indices distinguish the kind of the fields, $F.T.\,\{f(y,z)\}$ denotes the Fourier transformation for $f(y,z)$, and $|_{0}$ means taking $(b,f,{\overline f})=(b_{0},f_{0},{\overline f_{0}})$. ${\widetilde {\rm sTr}}$ is defined by
\begin{equation}
{\widetilde{\rm sTr}}\pmatrix{M&O\cr P&N\cr}={\widetilde{\rm Tr}}M-{\widetilde{\rm Tr}}N,
\end{equation}
where $M$ and $N$ are the matrices having Greek indices for bosons and fermions, respectively, and
\begin{equation}
{\widetilde {\rm Tr}}[M_{\alpha\beta}(p,q)]=\sum_{\alpha}\int^{\mu^{2}}_{\mu^{2}-\Delta \mu^{2}}d^{4}p~M_{\alpha\alpha}(p,p).\label{t-Trace}
\end{equation}
The integral on the right-hand side of Eq.~(\ref{t-Trace}) is defined with the Euclidean momentum ($p^{0}=ip^{4}$).

Now, we parameterize the Higgs doublet as
\begin{equation}
\Phi=\frac{1}{\sqrt{2}}\left(
\begin{array}{c}
\chi_{2}+i\chi_{1}\\
\phi -i\chi_{3}
\end{array}
\right),\label{parameterization}
\end{equation}
where $\chi_{i=1,2,3}$ are the would-be N.G.~bosons and $\phi$ is the physical Higgs boson. We focus on $\phi$ by setting the other fields equal to zero from now. Then Eq.~(\ref{1-loop}) can be written as
\begin{equation}
{\widetilde \Delta} S[\phi]=\int d^{4}x\left[-\Delta Z\frac{1}{2}\phi\partial^{2}\phi-\sum_{n=2,4,\cdots}^{\infty}{\widetilde \Delta}\lambda_{n}\phi^{n}-\sum_{m,n}^{\infty}{\widetilde \Delta}c_{mn}H(\phi^{n},\partial^{m})\right],
\end{equation}
where
\begin{equation}
\lambda_{2}=\frac{1}{2}m^{2},~~\lambda_{4}=\frac{1}{4}\lambda\label{m2 and lambda}
\end{equation}
and $H(\phi^{n},\partial^{m})$ denotes higher derivative terms.

Step II: One can consider that the momentum cutoff has decreased through Step I such as $\mu^{2} \to b^{2}\mu^{2}$, where $b^{2}=1-(\Delta \mu^{2}/\mu^{2})$. Here, we introduce new momentum and coordinate variables, $k'^{2}=b^{-2}k^{2}$ and $x'^{2}=b^{2}x^{2}$. Note that the cutoff for $k'^{2}$ is restored. Using $x'$, we can write as
\begin{eqnarray}
S[\phi]+{\widetilde \Delta} S[\phi]&=&\int d^{4}x'b^{-4}\Biggl[-\frac{1}{2}b^{2}(1+\Delta Z)\phi\partial^{'2}\phi -\sum_{n=2,4,\cdots}^{\infty}(\lambda+{\widetilde \Delta}\lambda_{n})\phi^{n}\nonumber \\
&&-\sum_{m,n}^{\infty}(c_{nm}+{\widetilde \Delta}c_{nm})b^{m}H(\phi^{n},\partial^{'m})\Biggr].\label{step 2}
\end{eqnarray}

Step III: We normalize the kinetic term on the right-hand side of Eq.~(\ref{step 2}) through the re-definition $\phi \to b(1+\Delta Z)^{-1/2}\phi$:
\begin{equation}
S[\phi]+{\widetilde \Delta} S[\phi] \to S[\phi]+\Delta S[\phi],
\end{equation}
where
\begin{equation}
\Delta S[\phi]=\int d^{4}x\left\{-\sum_{n=2,4,\cdots}^{\infty}\Delta\lambda_{n}\phi^{n}-\sum_{m,n}^{\infty}\Delta c_{mn}H(\phi^{n},\partial^{m})\right\}\label{step 3}
\end{equation}
and
\begin{eqnarray}
\Delta \lambda_{n}&=&{\widetilde \Delta}\lambda_{n}+\lambda_{n}~\frac{1}{2}\left\{(4-n)\frac{\Delta \mu^{2}}{\mu^{2}}-n\Delta Z\right\},\label{Delta}\\
\Delta c_{mn}&=&{\widetilde \Delta}c_{mn}+c_{mn}\frac{1}{2}~\left\{(4-m-n)\frac{\Delta \mu^{2}}{\mu^{2}}-n\Delta Z\right\}.
\end{eqnarray}
We define $\Delta S[\phi]$ as the variation of the action under the infinitesimal renormalization transformation in our cutoff model.

The variation of the full action under the infinitesimal renormalization transformation will not be gauge invariant, and so the effective lagrangian density will not be gauge invariant in general. This is the problem of the gauge invariance in our cutoff model mentioned above and \S 1.

The $\beta$ functions for $m^{2}$ and $\lambda$ are
\begin{equation}
\beta_{m^{2}}=-2\mu^{2}\frac{\Delta m^{2}}{\Delta \mu^{2}},~~~\beta_{\lambda}=-2\mu^{2}\frac{\Delta \lambda}{\Delta \mu^{2}},
\end{equation}
where we have identified $-\Delta/\Delta \mu^{2}$ with $d/d \mu^{2}$ because $\Delta S[\phi]$ is the variation with decreasing $\mu$. In the 't Hooft-Landau gauge, we have
\begin{eqnarray}
\beta_{m^{2}}&=&\frac{1}{4\pi}\left\{\frac{-12\alpha_{\lambda}\mu^{4}}{\mu^{2}+m^{2}}+2m^{2}4\pi(\widetilde{\gamma}-1)+\left(12\alpha_{t}-\frac{9}{2}\alpha_{2}-\frac{9}{10}\alpha_{1}\right)\mu^{2}\right\},\label{bm}\\
\beta_{\lambda}&=&\frac{24\alpha_{\lambda}^{2}\mu^{4}}{(\mu^{2}+m^{2})^{2}}+\frac{27}{200}\alpha_{1}^{2}+\frac{9}{20}\alpha_{1}\alpha_{2}+\frac{9}{8}\alpha_{2}^{2}-6\alpha_{t}^{2}+4\alpha_{\lambda}4\pi\widetilde{\gamma}~=4\pi\beta_{\alpha_{\lambda}},~~~~~~~~~~\label{bl}
\end{eqnarray}
where
\begin{equation}
\widetilde{\gamma}=\mu^{2}\frac{\Delta Z}{\Delta \mu^{2}}=\frac{1}{4\pi}\left\{3\alpha_{t}-\frac{\mu^{2}}{\mu^{2}+m^{2}}\left(\frac{9}{20}\alpha_{1}+\frac{9}{4}\alpha_{2}\right)\right\}.
\end{equation}
We have calculated the $\beta$ functions neglecting higher dimensional terms in the effective action on the assumption that their contributions will be $o(\mu^{2}/m_{Pl}^{2})$. Because the calculations are not reliable for $\mu \sim m_{Pl}$, our discussion based on the $\beta$ functions is not reliable for $\Lambda \sim m_{Pl}$. However, we consider that $\Lambda$ is not so large compared to $m_{Z}$ because a fine-tuning will be required if $\Lambda \gg m_{Z}$ as we will see in the next paragraph.

Let us try to determine the running of $\lambda$ and $m^{2}$ using Eqs.~(\ref{bm}) and (\ref{bl}). Because $\beta_{\lambda}$ contains $m^{2}$ in the expression, we must solve the differential equations for $\lambda$ and $m^{2}$ simultaneously. To solve them numerically, a boundary value for $m^{2}$ at $\Lambda$ is necessary because the boundary value for $\lambda$ is given at $\Lambda$, i.e., Eq.~(\ref{composite condition2-1}). Practically, we solve the differential equations with tuning the value of $m^{2}(\Lambda)$ so as to obtain the solutions for $\lambda(\mu)$ and $m^{2}(\mu)$ that satisfy the condition for the minimum,
\begin{equation}
-\frac{m^{2}(m_{Z})}{\lambda(m_{Z})}=v^{2},\label{extremal}
\end{equation}
where $v$ is of Eq.~(\ref{v})\footnote{The value $v=246.22$ GeV should be consider to be one at $\mu=m_{W}$ because it is determined by the experiment of the muon decay~\cite{vev}. According to Ref.~\cite{vev}, we consider $v(m_{Z})$ to be equal to $v(m_{W})$ approximately.}. When $\Lambda \gg m_{Z}$, the tuning for $m^{2}(\Lambda)$ becomes hard. This is the problem of fine-tuning mentioned above and in \S 4. Here, we consider the case $m^{2}\ll \mu^{2}$. In the leading order of the $m^{2}/\mu^{2}$ expansion, we have
\begin{eqnarray}
\beta_{m^{2}}&=&\beta_{m^2}^{\mbox{{\scriptsize MS}}}-2m^{2}+\frac{1}{4\pi}\left(12\alpha_{t}-12\alpha_{\lambda}-\frac{9}{10}\alpha_{1}-\frac{9}{2}\alpha_{2}\right)\mu^2,\\
\beta_{\lambda}&=&4\pi\beta_{\alpha_\lambda}^{\mbox{{\scriptsize MS}}},\label{approx-beta}
\end{eqnarray}
where $\beta_{\alpha_\lambda}^{\mbox{{\scriptsize MS}}}$ is of Eq.~(\ref{beta-a-lambda}), and
\begin{equation}
\beta_{m^{2}}^{\mbox{\scriptsize MS}}=\frac{2m^{2}}{4\pi}\left(6\alpha_{\lambda}+3\alpha_{t}-\frac{9}{20}\alpha_{1}-\frac{9}{4}\alpha_{2}\right).
\end{equation}
In this case we can determine $\lambda$ independently of $m^2$ as performed in \S 3 because $\beta_{\alpha_\lambda}^{\mbox{{\scriptsize MS}}}$ does not contain $m^2$. Then, we can obtain a value of $m^{2}(m_{Z})$ from Eq.~(\ref{extremal}) without any fine-tuning, which can be used as a boundary value to determine the running of $m^{2}(\mu)$.

If we use $\beta_{m^{2}}^{\mbox{\scriptsize MS}}$ with Eq.~(\ref{extremal}), we can see that $m^{2}$ is negative in $\Lambda_{{\rm QCD}}<\mu \le m_{Pl}$. This means that if we adopt the ${\overline {\rm MS}}$  scheme in our scenario for the composite Higgs particle, the spontaneous symmetry breaking is not properly described as mentioned in \S 1.

\begin{figure}[t]
\begin{minipage}[t]{6.8cm}
 \leavevmode
 \epsfxsize= 68mm
 \epsfysize= 47mm
 \centerline{\epsfbox{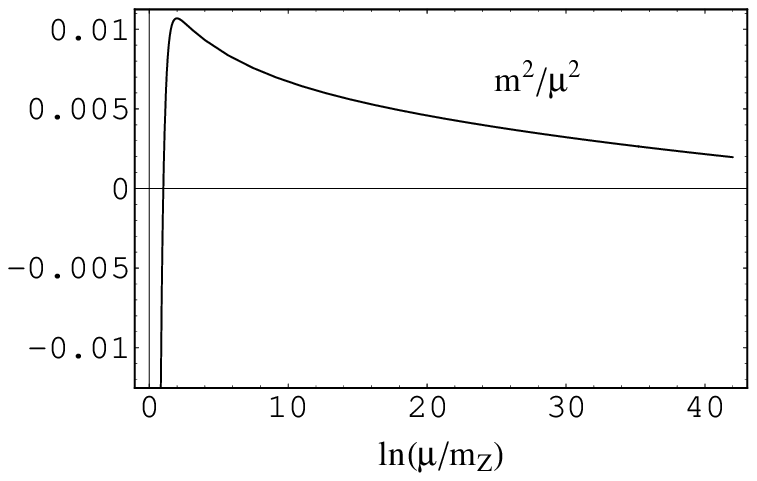}}
 \caption{The RG evolution of $m^{2}/\mu^{2}$ for $\Lambda=10^{4}$ GeV and $m_{t}=173.8$ GeV.}
 \label{zmm4}
\end{minipage}
\hspace{1mm}
\begin{minipage}[t]{6.8cm}
 \leavevmode
 \epsfxsize= 68mm
 \epsfysize= 47mm
 \centerline{\epsfbox{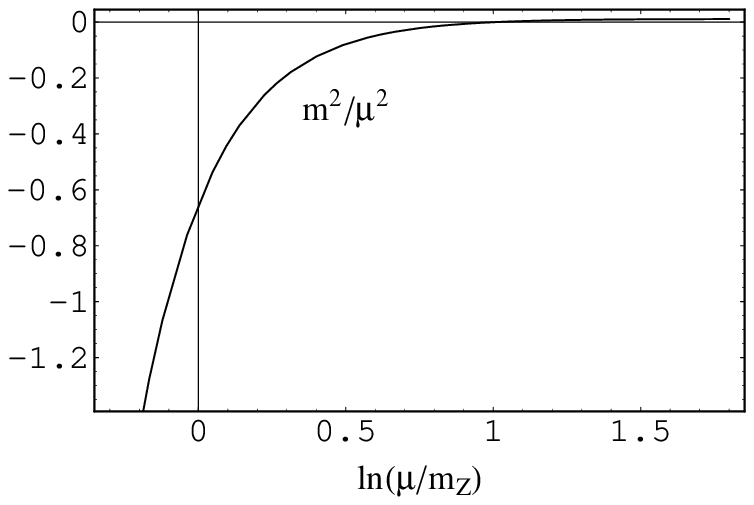}}
 \caption{$m^{2}/\mu^{2}$ in low energies. $\Lambda$ and $m_{t}$ are the same with ones of Fig.~3.}
 \label{zmm4R}
\end{minipage}
\end{figure}

Finally, we check the consistency of the use of $m^{2}\ll \mu^{2}$. As Figs.~\ref{zmm4} and \ref{zmm4R} show, $m^{2}/\mu^{2}$ is sufficiently small until the blowing down in the low-energy region appears.
We can see that $m^{2}/\mu^{2}<1$ in $0\le \mu \le m_{Pl}$.

\section{A form for the effective lagrangian}
We have discussed the vacuum stability based on the "tree potential" $V^{(0)}(t)=\{m^{2}(t)/2\}\phi^{2}+\{\lambda(t)/4\}\phi^{4}$ of the Higgs field assuming that higher dimensional terms can be neglected for $\phi^2 \to \infty$. In this Appendix, we give a desired form for the effective lagrangian to make our assumption valid.

In the ordinary discussion of the vacuum stability, the Coleman-Weinberg potential is used with the RG improvement~\cite{stability bound-2,HKKN,stability bound-1}. In our discussion, however, we cannot use it because we are treating our model as a cutoff model with having Wilson's renormalization approach in mind. The effective potential $V(t,\phi)$ of our model will be obtained by solving the RG equation for the effective lagrangian. From Eq.~(\ref{step 3}), we have
\begin{equation}
\Delta V(t,\phi)=\sum_{n=2,4,\cdots}^{\infty}\Delta\lambda_{n}\phi^{n}\label{RG eq-0}
\end{equation}
 by taking $\phi=$ constant, where $\Delta V(t,\phi)$ is defined by
\begin{equation}
\Delta V(t,\phi)\int d^{4}x=-\Delta S[\phi={\rm constant}].
\end{equation}
Using $-2\mu^{2}\Delta V(t,\phi)/\Delta \mu^{2}=\partial V(t,\phi)/\partial t$, we can reduce Eq.~(\ref{RG eq-0}) to
\begin{equation}
\frac{\partial V}{\partial t}=-4V+(1+{\widetilde \gamma})\phi \frac{\partial V}{\partial \phi}+T(t,\phi),\label{RGE for effective potential}
\end{equation}
where $T(t,\phi)$ is defined by
\begin{eqnarray}
&&T(t,\phi)\frac{\Delta \mu^{2}}{2\mu^{2}}\cdot\int d^{4}x \nonumber \\
&&=\frac{i}{2}{\widetilde {\rm Tr}}\,{\rm Ln}\left[\frac{1}{2} F.T.\!\left\{\left.\frac{\delta^{2}S}{\delta b_{\alpha}(y) \delta b_{\beta}(z)}~\right|_{0}\right\}\right]-i{\widetilde {\rm Tr}}\,{\rm Ln}\left[F.T.\!\left\{\left.\frac{\delta^{2}S}{\delta f_{\alpha}(y) \delta {\overline f_{\beta}}(z)}~\right|_{0}\right\}\right]~~~~~~~~~~
\end{eqnarray}
and $|_{0}$ means taking $\phi=$ constant and $b_{\alpha}=f_{\alpha}={\overline f_{\alpha}}=0$ for the other fields. We cannot determine $V(t,\phi)$ only from this equation because it is necessary to know the full effective lagrangian ${\cal L}$ to know $T(t,\phi)$. From now, we give asymptotic solutions of the equation for $\phi^{2} \to \infty$ with imposing artificial conditions on $T(t,\phi)$ or the effective lagrangian. We will consider three cases, and the last case is the desired one.

We first examine the case,
\begin{equation}
\lim_{\phi^{2} \to \infty}\frac{T(t,\phi)}{\phi^{2}}=0.
\end{equation}
By putting
\begin{equation}
V(t,\phi)=V^{(0)}(t,\phi)+v(t,\phi),~~~v(t,\phi)=\phi^{2}\omega(t,\phi),
\end{equation}
we have
\begin{equation}
\frac{\partial \omega}{\partial t}-\left(1+{\widetilde \gamma}\right)\phi \frac{\partial \omega}{\partial \phi}+2\left(1-{\widetilde \gamma}\right)\omega=-\frac{1}{2}\left\{\beta_{m^{2}}+2m^{2}\left(1-{\widetilde \gamma}\right)\right\}-\frac{1}{4}\left(\beta_{\lambda}-4{\widetilde \gamma}\lambda\right)\phi^{2}
\end{equation}
from Eq.~(\ref{RGE for effective potential}) for $\phi^{2} \to \infty$. The general solution of this equation yields
\begin{equation}
v(t,\phi)=-\frac{1}{2}m^{2}\phi^{2}-\frac{1}{4}\lambda\phi^{4}+{\rm e}^{-4t}F\left({\rm e}^{\int_{0}^{t}dt\left(1+{\widetilde \gamma}\right)}|\phi|\right),\label{trivial}
\end{equation}
where $F$ is a function. Note that the first two terms on the left-hand side of Eq.~(\ref{trivial}) cancel $V^{(0)}$. The functional form for $F$ is determined from the condition that $V$ becomes quadratic at $\Lambda$, i.e., Eq.~(\ref{compositeness0-3}), and we have
\begin{equation}
V(t,\phi)=C~{\rm e}^{-2\int_{0}^{t}dt\left(1-{\widetilde \gamma}\right)}\phi^{2},
\end{equation}
where $C$ is a constant having mass dimension $2$. This means that our scenario for the composite Higgs particle does not hold in this case.

Next, we consider the case,
\begin{equation}
\lim_{\phi^{2} \to \infty}\frac{T(t,\phi)}{\phi^{2}}\ne 0.\label{ne 0}
\end{equation}
First in this case, we examine the following candidate for the effective lagrangian\footnote{We can also add to ${\cal L}$ other higher dimensional terms that give no effect on $\lim_{\phi^{2} \to \infty}T(t,\phi)/\phi^{2}$ and $v(t,\phi)$.},
\begin{equation}
{\cal L}={\cal L}^{(0)}-f(t){\rm e}^{x(t)|\Phi|^{2}+y(t)(|\Phi|^{2})^{2}}-u(t,|\Phi|^{2}),\label{desired-1}
\end{equation}
where ${\cal L}^{(0)}$ is the ordinary renormalizable lagrangian of the standard model, and $u(t,|\Phi|^{2})$ denotes higher dimensional terms that have no effect on $\lim_{\phi^{2} \to \infty}T(t,\phi)/\phi^{2}$. Functional forms for $x(t)>0$, $y(t)>0$ and $f(t)>0$ are determined from now. The effective lagrangian yields
\begin{equation}
\lim_{\phi^{2} \to \infty}\frac{T(t,\phi)}{\phi^{2}}=-\frac{\mu^{4}}{4\pi^{2}}\left\{\frac{x(t)}{2}+\frac{y(t)}{4}\phi^{2}\right\},~~~(\mu=m_{Z}{\rm e}^{t})
\end{equation}
and then Eq.~(\ref{RGE for effective potential}) reduces to
\begin{eqnarray}
&&\frac{\partial \omega}{\partial t}-\left(1+{\widetilde \gamma}\right)\phi \frac{\partial \omega}{\partial \phi}+2\left(1-{\widetilde \gamma}\right)\omega \nonumber \\
&&=-\frac{1}{2}\left\{\beta_{m^{2}}+2m^{2}\left(1-{\widetilde \gamma}\right)+\frac{\mu^{4}x(t)}{4\pi^{2}}\right\}-\frac{1}{4}\left\{\beta_{\lambda}-4{\widetilde \gamma}\lambda+\frac{\mu^{4}y(t)}{4\pi^{2}}\right\}\phi^{2}~~~~~~~~~~~~
\end{eqnarray}
for $\phi^{2} \to \infty$. The general solution of this equation yields
\begin{eqnarray}
v(t,\phi)&=&-\frac{1}{2}\left\{m^{2}+\frac{1}{4\pi^{2}}{\rm e}^{-2\int_{a}^{t}dt\left(1-{\widetilde \gamma}\right)}\int_{b}^{t}dt~\mu^{4}x(t)~{\rm e}^{2\int_{a}^{t}dt\left(1-{\widetilde \gamma}\right)}\right\}\phi^{2}\nonumber \\
&&-\frac{1}{4}\left\{\lambda +\frac{1}{4\pi^{2}}{\rm e}^{4\int_{a}^{t}dt{\widetilde \gamma}}\int_{b}^{t}dt~\mu^{4}y(t)~{\rm e}^{-4\int_{a}^{t}dt{\widetilde \gamma}}\right\}\phi^{4}\nonumber \\
&&+{\rm e}^{-4t}F\left({\rm e}^{\int_{0}^{t}dt\left(1+{\widetilde \gamma}\right)}|\phi|\right),\label{result of desired-1}
\end{eqnarray}
where $a$ and $b$ are constants and $F$ is a function. On the other hand, Eq.~(\ref{desired-1}) requires
\begin{equation}
v(t,\phi)=f(t){\rm e}^{x(t)\phi^{2}/2+y(t)\phi^{4}/4}+u(\phi^{2}/2).\label{first possibility}
\end{equation}
By comparing this with Eq.~(\ref{result of desired-1}), we have
\begin{eqnarray}
x(t)&=&C_{1}~{\rm e}^{2\int_{0}^{t}dt\left(1+{\widetilde \gamma}\right)},\\
y(t)&=&C_{2}~{\rm e}^{4\int_{0}^{t}dt\left(1+{\widetilde \gamma}\right)},\\
f(t)&=&f~{\rm e}^{-4t},
\end{eqnarray}
where $C_{1,2}>0$ and $f>0$ are constants having mass dimensions $-2$, $-4$ and $4$,  respectively\footnote{Equation (\ref{result of desired-1}) also yields certain restrictions on the functional form for $u(t,|\Phi|^{2})$. Under this restrictions we can choose the functional form for $u(t,|\Phi|^{2})$ in consistent with the assumption that it has no effect on $\lim_{\phi^{2} \to \infty}T(t,\phi)/\phi^{2}$.}. Note that $f(t)$ does not vanish at any finite scale, and the first term on the right-hand side of Eq.~(\ref{first possibility}) dominates in $V(t,\phi)$ for $\phi^{2} \to \infty$. Thus, $\lambda \ge 0$ is not the condition for the vacuum stability in this case. This means that our scenario for the composite Higgs particle does not hold in this case too.

Finally, we consider the case that Eq.~(\ref{ne 0}) holds thanks to derivative coupling terms. As a candidate, we examine
\begin{eqnarray}
{\cal L}&=&{\cal L}^{(0)}+f_{1}(t)~{\rm e}^{x_{1}(t)|\Phi|^{2}+y_{1}(t)(|\Phi|^{2})^{2}}|D_{\mu}\Phi|^{2}\nonumber \\
&&+f_{2}(t)~{\rm e}^{x_{2}(t)|\Phi|^{2}+y_{2}(t)(|\Phi|^{2})^{2}}~{\overline \Psi_{t}}{\ooalign{\hfil/\hfil\crcr $D$}}^{f}\Psi_{t}+{\cal L}^{{\rm H}},\label{second candidate}
\end{eqnarray}
where ${\ooalign{\hfil/\hfil\crcr $D$}}^{f}$ is the covariant derivative for the top field $\Psi_{t}$, $x_{1,2}(t) \ge 0,~y_{1,2}(t) \ge 0$ and $f_{1,2}(t)>0$ are certain functions of $t$, and ${\cal L}^{{\rm H}}$ denotes higher dimensional terms that have no effect on $\lim_{\phi^{2} \to \infty}T(t,\phi)/\phi^{2}$. We give certain restrictions on $x_{1,2}(t)$, $y_{1,2}(t)$ and $f_{1,2}(t)$ from now. The effective lagrangian yields
\begin{equation}
\lim_{\phi^{2} \to \infty}\frac{T(t,\phi)}{\phi^{2}}=\frac{\mu^{4}}{(4\pi)^{2}}\left\{\frac{1}{2}(12x_{2}-13x_{1})+\frac{1}{4}(12y_{2}-13y_{1})\phi^{2}\right\}.\label{xy}
\end{equation}
If $x_{1,2}(t)$ and $y_{1,2}(t)$ satisfy the conditions
\begin{eqnarray}
12x_{2}(t)-13x_{1}(t)&=&-\frac{(4\pi)^{2}}{\mu^{4}}\left\{\beta_{m^{2}}+2m^{2}\left(1-{\widetilde \gamma}\right)\right\},\label{x12}\\
12y_{2}(t)-13y_{1}(t)&=&-\frac{(4\pi)^{2}}{\mu^{4}}\left(\beta_{\lambda}-4{\widetilde \gamma}\lambda\right),\label{y12}
\end{eqnarray}
Eq.~(\ref{RGE for effective potential}) reduces to
\begin{equation}
\frac{\partial \omega}{\partial t}-\left(1+{\widetilde \gamma}\right)\phi \frac{\partial \omega}{\partial \phi}+2\left(1-{\widetilde \gamma}\right)\omega =0
\end{equation}
for $\phi^{2} \to \infty$. The general solution of this equation yields
\begin{equation}
v(t,\phi)={\rm e}^{-4t}F\left({\rm e}^{\int_{0}^{t}dt\left(1+{\widetilde \gamma}\right)}|\phi|\right),
\end{equation}
where $F$ is a function. From the condition that $V(t,\phi)$ becomes quadratic at $\Lambda$, we have $v(t,\phi)=0$. After all, if ${\cal L}$ of Eq.~(\ref{second candidate}) is realized with the conditions (\ref{x12}) and (\ref{y12}), the "tree potential" is an asymptotic solution of the RG equation (\ref{RGE for effective potential}) for $\phi^{2} \to \infty$. In this case our discussion on the vacuum stability based on the "tree potential" is valid, i.e., $\lambda \ge 0$ is the condition for the vacuum stability.

In our scenario for the composite Higgs particle, ${\cal L}$ itself must become quadratic with respective to $\phi$ at $\Lambda$. This requirement implies that $x_{1,2}(t)$ and $y_{1,2}(t)$ vanish at $\Lambda$, and so we obtain
\begin{eqnarray}
\beta_{m^{2}}(\Lambda)&=&2m^{2}(\Lambda)\left\{{\widetilde \gamma}(\Lambda)-1\right\},\\
\beta_{\lambda}(\Lambda)&=&0
\end{eqnarray}
from Eqs.~(\ref{x12}), (\ref{y12}) and (\ref{composite condition2-1}). The conditions for the $\beta$ functions should be considered with taking account of the contributions from higher dimensional terms in ${\cal L}$ to the $\beta$ functions that were neglected approximately\footnote{To make the approximation valid, $f(t)\sim o(\mu^{2}/m_{Pl}^{2})$ is required.} when we gave Eqs.~(\ref{bm}) and (\ref{bl}). The conditions will give some restriction on the value of $\Lambda$ if they can hold. The further investigation of the conditions remains as future works.

\end{document}